\documentclass[onecolumn,preprint,showpacs,preprintnumbers,amsmath,amssymb]{revtex4}
\usepackage{graphicx}
\usepackage{epsfig}
\usepackage{dcolumn}
\usepackage{bm}
\usepackage{color}

\def\Journal#1&#2&#3(#4){#1{\bf #2}, #3 (#4)}

\def\PRL{Phys.  Rev.  Lett.  }
\def\PRD{Phys.  Rev.  { D }}

\def\etal{{\it et al.}}

\def\bec{\begin{center}}
\def\eec{\end{center}}

\def\g {\gamma}
\def\chiscom {\chi^2_{com}}
\def \ppb {p\bar p}
\def \ks {K_S^0}

\def \chis {\chi^2}
\newcommand{\BR}{\mathcal{B}}
\newcommand{\rar}{\rightarrow}
\newcommand{\tppkk}{2(\pi^+\pi^-)K^+K^-}
\newcommand{\psip}{\psi(2S)}
\newcommand{\psipto}{\psip\rightarrow}
\newcommand{\jpsi}{J/\psi}
\newcommand{\jpsito}{J/\psi\rightarrow}
\newcommand{\pipi}{\pi^+\pi^-}
\newcommand{\ppbar}{p\bar{p}}

\newcommand{\kskp}{K^0_S K^+ \pi^- + c.c.}
\newcommand{\kstaro}{K^{\ast0}}
\newcommand{\kstarb}{\bar{K}^{\ast0}}

\newcommand{\gpppr}{\gamma \pi^+\pi^-p\bar{p}}
\newcommand{\gkkkk}{\gamma K^+K^-K^+K^-}
\newcommand{\pp}{\pi^+\pi^-}

\newcommand{\MC}{~\hbox{MeV}/c^2}

\newcommand{\GC}{~\hbox{GeV}/c^2}
\newcommand{\piz}{\pi^0}
\newcommand{\pip}{\pi^+}
\newcommand{\pim}{\pi^-}
\newcommand{\fptk}{2(\pp)\kk}
\newcommand{\kk}{K^+K^-}

\begin{document}
\title{\boldmath Measurement of $\psip$ Radiative Decays }
\author{
M.~Ablikim$^{1}$,              J.~Z.~Bai$^{1}$,               Y.~Ban$^{12}$,
X.~Cai$^{1}$,                  H.~F.~Chen$^{17}$,
H.~S.~Chen$^{1}$,              H.~X.~Chen$^{1}$,              J.~C.~Chen$^{1}$,
Jin~Chen$^{1}$,                Y.~B.~Chen$^{1}$
% ...
\\
\vspace{0.2cm}
(BES Collaboration)\\
\vspace{0.2cm} {\it
$^{1}$ Institute of High Energy Physics, Beijing 100049, People's Republic of China\\
% ...
$^{12}$ Peking University, Beijing 100871, People's Republic of China\\
% ...
$^{17}$ University of Science and Technology of China, Hefei 230026, People's Republic of China\\
% ...
}}

\begin{abstract}

Using 14 million $\psi(2S)$ events accumulated at the BESII
detector, we report first measurements of branching fractions or
upper limits for $\psip$ decays into $\gamma\ppbar$, $\gamma
2(\pipi)$, $\gamma \kskp$, $\gamma K^+ K^- \pipi$, $\gamma\kstaro
K^- \pi^+ +c.c.$, $\gamma K^{*0}\bar K^{*0}$, $\gamma\pipi\ppbar$,
$\g2(\kk)$, $\gamma3(\pp)$, and $\gamma\tppkk$ with the invariant
mass of hadrons below 2.9$\GC$. We also report branching fractions
of $\psip$ decays into $2(\pp)\piz$, $\omega\pp$, $\omega
f_2(1270)$, $b_1^\pm\pi^\mp$, and $\pi^0\fptk$.

\end{abstract}


\maketitle

Besides the conventional meson and baryon states, QCD also
predicts a rich spectrum of glueballs, meson hybrids, and
multi-quark states in the 1.0 to 2.5~$\hbox{GeV}/c^2$ mass region.
Therefore, searches for the evidence of these exotic states play
an important role to test QCD. Such studies have been performed in
$\jpsi$ radiative decays for a long time~\cite{Jdecay, QWG}, while
studies in $\psip$ radiative decays have been limited due to low
statistics in previous experiments~\cite{PDG, QWG}. The radiative
decays of $\psip$ to light hadrons are expected to
contribute about 1\% to the total $\psip$ decay
width~\cite{PRD-wangp}. However, the measured channels only sum up
to about 0.05\%~\cite{PDG}.

In this Letter, we present first measurements of  $\psip$ decays
into $\gamma\ppbar$, $\gamma 2(\pipi)$, $\gamma \kskp$, $\gamma
K^+ K^- \pipi$, $\gamma\kstaro K^- \pi^+ +c.c.$, $\gamma
K^{*0}\bar K^{*0}$, $\gamma\pipi\ppbar$, $\g2(\kk)$,
$\gamma3(\pp)$, and $\gamma\tppkk$,  with the invariant mass of
the hadrons ($m_{hs}$) less than 2.9$\GC$ for each decay mode.
Measurements of $\psip$ decays into $\pi^02(\pipi)$ and
$\pi^02(\pipi)K^+K^-$ are also presented and are used for
estimating backgrounds contributing to $\psip$ decays into $\gamma
2(\pipi)$ and $\gamma\tppkk$, respectively.

The data samples used in this analysis consist of
$(14.00\pm0.56)\times 10^6$ $\psip$ events
($\mathcal{L}=19.72$~pb$^{-1}$) and 6.42 pb$^{-1}$ of continuum data
at $\sqrt{s}=3.65$~GeV, acquired with the BESII detector. BESII is a
conventional solenoid magnetic detector~\cite{BES-II}, which
consists of a vertex chamber (VC), a main drift chamber (MDC), a
time-of-flight (TOF) system, a barrel shower counter (BSC), and a
muon counter. MDC also measures the energy loss ($dE/dx$) for
particle identification. A GEANT3 based Monte Carlo (MC)
program~\cite{SIMBES} is used for the simulation.

A common set of requirements is used to select charged tracks and
photon candidates for all channels. Each charged track is required to
be well fitted to a helix in the MDC, to be within the polar angle
region $|\cos\theta|<0.8$, and to have a transverse momentum larger
than 70~MeV/$c$. The total charge of the good charged tracks in each
event is required to be zero.  Each photon candidate is required to
have an energy deposit in the BSC greater than 50~MeV, to be isolated
from charged tracks by more than $15^{\circ}$, and to have the angle
between the cluster development direction in the BSC and the photon
emission direction less than $37^{\circ}$.

For each decay mode, the number of charged tracks is required to
be equal to the number of charged stable hadrons in the
corresponding final state. The TOF and $dE/dx$ measurements of the
charged track are used to calculate $\chi_{\textrm{PID}}^2$ values
and the corresponding confidence levels (C.L.) for the hypotheses
that the particle is a pion, kaon, or proton. All charged tracks
in the selection of $\psip\to \g\ppb$, $\g2(\pp)$, and $\g2(\kk)$
are required to be consistent with the proton, pion, or
kaon assumption with the corresponding C.L. greater than 1\%. For
$\psip\to \g\kk\pp$, $\g\pp\ppb$, and $\g2(\pp)\kk$, only two
charged tracks are required to be identified as kaons or protons,
respectively.

Next, the selected charged tracks and the photon with the largest
energy are fitted kinematically using energy and momentum
conservation constraints (4C), and the combined probability,
$prob(\chi^2_{com}, ndf)$ is required to be greater than 1\%, where
$ndf$ is the number of degrees of freedom and $\chi^2_{com}$ is the
sum of the $\chi^2$ of the kinematic fit ($\chi^2_{4C}$) and
particle identification ($\chi_{\textrm{PID}}^2(i)$), {\it i.e.}
$\chi^2_{com}=\sum\limits_{i} \chi_{\textrm{PID}}^2(i) +
\chi_{4C}^2$, where $i$ runs over all charged tracks. For $\psip\to
\piz2(\pp)$ and $\piz 2(\pp)\kk$, if there are more than two photons
in an event, the photon-pair with the minimum $\chiscom$ is chosen.
To remove background from charged particle misidentification, the
$\chi^2_{com}$ for the signal hypothesis is required to be less than
those for background.

To select $\g\ks K^+\pim+c.c.$ events, the $\ks$ candidate must have
a decay length in the transverse plane greater than 0.5~cm. In
selecting $\psip\to\g\kk\pp$, contaminations from $\psip\to\g\ks
K\pi$ is removed by requiring the invariant mass of the two pions to
be outside of the $\ks$ mass region, $i.e.$,
$|m_{\pi\pi}-m_{\ks}|>0.04$~GeV/$c^2$.

To reject $\psip$ transitions into other charmonium states, $m_{hs}$
is required to be less than 2.9~GeV/$c^2$ for each decay mode. If
there is possible background from $\psipto \pipi \jpsi$, it is
removed by requiring
$|m^{\pi^{+}\pi^{-}}_{recoil}-m_{J/\psi}|>0.05$~GeV/$c^2$, where
$m^{\pi^{+}\pi^{-}}_{recoil}$ is the mass recoiling from each
possible $\pipi$ pair.

\begin{figure}[h] \centering
\includegraphics[height=10cm, width=0.45\textwidth]{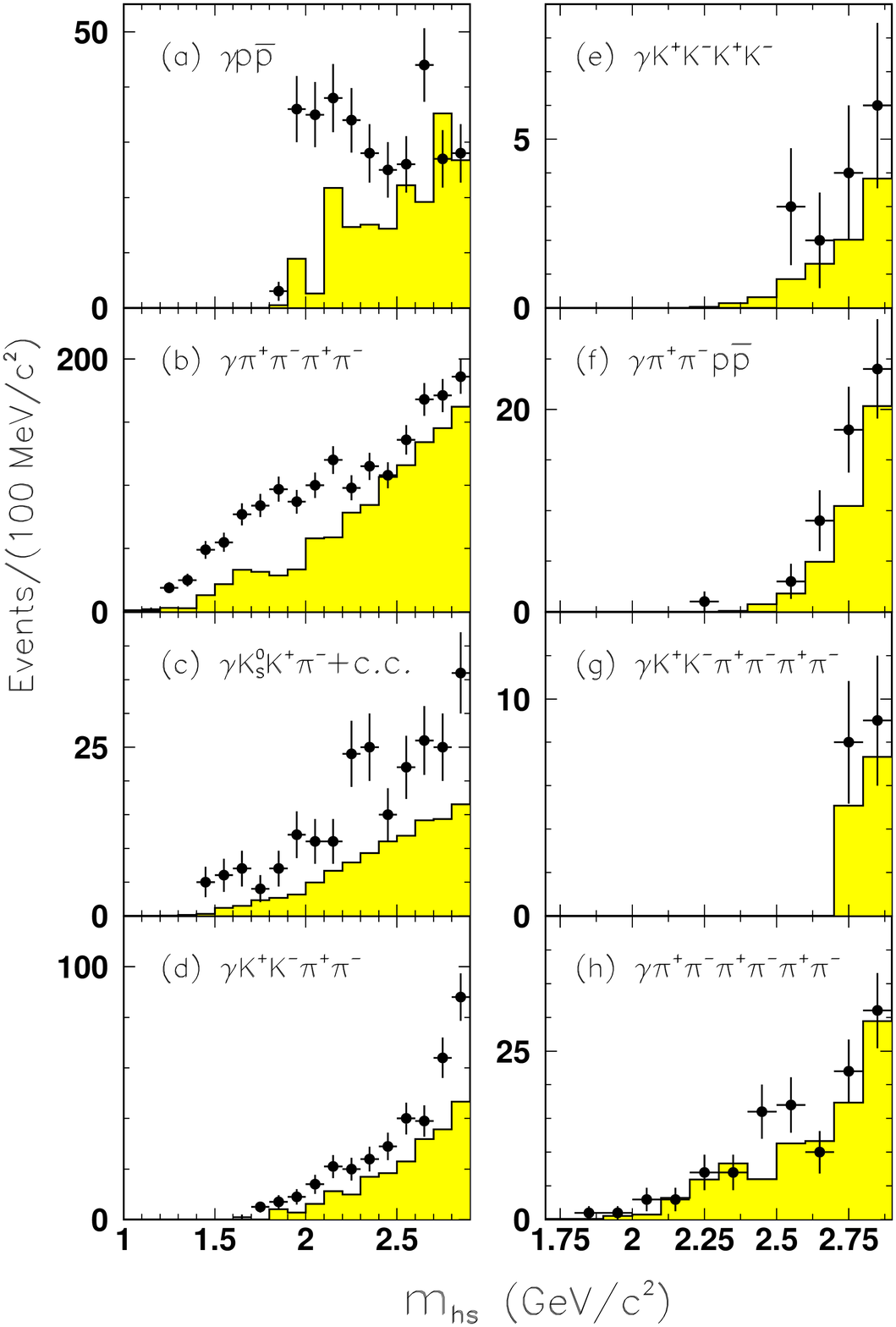}
\caption{ \label{spectrum} Invariant mass distributions of the hadrons
in each final state (dots with error bars). The shaded histograms
are backgrounds. }
\end{figure}

\begin{figure}[h] \centering
\includegraphics[width=0.45\textwidth]{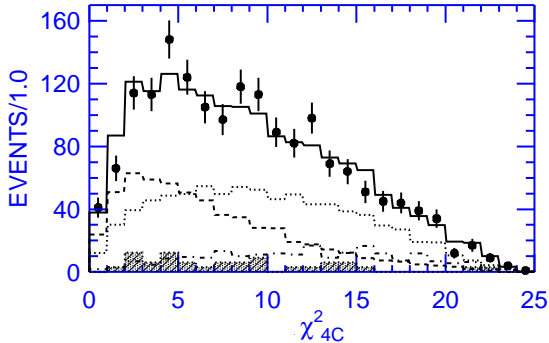}
\caption{ \label{chisqfit} The fitted $\chi^2_{4C}$ distribution for
$\psipto\gamma 2(\pip\pim)$ candidate events. The dots with error bars
are data. The solid line is the fitted result with the four
components: signal events (dashed line), MC simulated multi-photon
backgrounds (dotted line), QED processes (hatched histogram), and the
other backgrounds (dot-dashed line). }
\end{figure}

Figure~\ref{spectrum} shows invariant mass distributions of the
hadrons for $\psip\to\g\ppb$, $\g2(\pp)$, $\g K_S^0K^+\pi^-+c.c.$,
$\g\kk\pp$, $\g2(\kk)$, $\g\pp\ppb$, $\g2(\pp)\kk$, and $\g3(\pp)$
below 2.9~GeV/$c^2$, where backgrounds are shown as shaded
histograms. The backgrounds of each decay mode fall into three
classes: QED processes, estimated using the continuum data;
multi-photon backgrounds, e.g. $\psip\to \pi^0+hadrons$,
$3\gamma+hadrons$, etc., where the $hadrons$ have the same charged tracks as the signal final state, estimated with the MC
simulation and normalized according to their branching
fractions~\cite{PDG,pppi0,2k3pi}; and other backgrounds, estimated
using the inclusive $\psip$ decay MC sample~\cite{lund}. The results show that the multi-photon backgrounds are dominant;
the QED background, and the other backgrounds,
including contamination between studied channels are lower. The observed $\chi^2_{4C}$ distributions include both
signal events and these backgrounds (see Fig.~\ref{chisqfit}).

The number of signal events is extracted by fitting the
observed $\chi^2_{4C}$ distributions with those of the signal and
background channels~\cite{fitchis}, {\it i.e.}
$\chis_{obs}=w_s\chis_{sig}+\sum_{w_{b_i}}w_{b_i}\chis_{bg}$,
where $w_s$ and $w_{b_i}$ are the weights of the signal and the
background decays, respectively. As an example,
Fig.~\ref{chisqfit} shows the $\chi^2$ distribution observed for
$\psipto\gamma 2(\pipi)$, together with the fitted $\chi^2$
distributions for the signal, multi-photon, QED, and other
background channels. In the fit, the weights
of the multi-photon backgrounds and the QED backgrounds ($w_b$)
are fixed to be the normalization factors, but the
weight of the signal ($w_s$) and the weight of the other backgrounds ($w_b$)
are free. With
this method, the numbers of signal events are extracted for each
decay mode with $m_{hs}<2.9$~GeV/$c^2$ and are listed in
Table~\ref{Tot-nev}.

In Fig.~\ref{spectrum}(a) there is an excess of events between
$p\bar{p}$ threshold and 2.5~GeV/$c^2$, but no significant narrow
structure due to the $X(1859)$ observed in $\jpsito\gamma
p\bar{p}$~\cite{jpsi-gpp}. A fit of the mass spectrum with an
acceptance-weighted $S$-wave Breit-Wigner for the $X$ resonance
(with mass and width fixed to 1859$\MC$ and 30$\MC$,
respectively), together with MC simulated background channels
along with $\psip\to\gamma\ppbar$ phase space
background~\cite{massres} yields $11.7\pm 6.7$ events with a
statistical significance of 2.0$\sigma$. The upper limit on the
branching fraction is determined to be $\mathcal B[\psipto\gamma
X(1859)\rar \gamma \ppbar]<5.4\times 10^{-6}$ at the 90\% C.L.

\begin{figure}\centering
\includegraphics[width=0.45\textwidth, height=10.0cm]{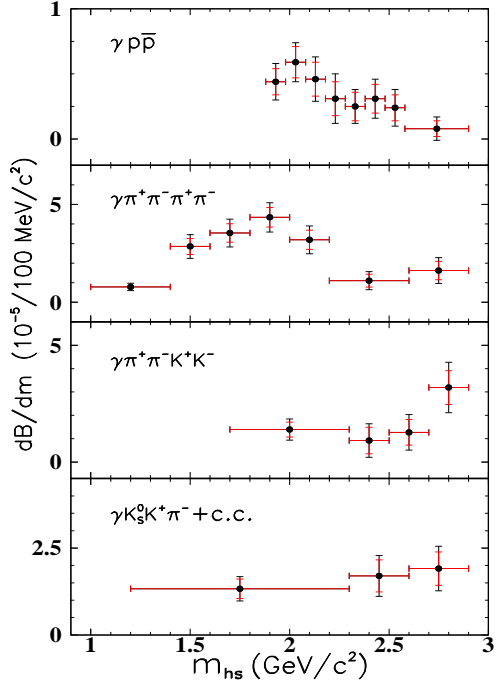}
\caption{ \label{diffbr} Differential branching fractions for $\psip$
decays into $\gamma\ppbar$, $\gamma 2(\pipi)$, $\gamma K^+ K^- \pipi$,
and $\gamma \kskp$ Here $m_{hs}$ is the invariant mass of the hadrons
in each final state. For each point, the smaller vertical error is
the statistical error, while the bigger one is the sum of statistical
and systematic errors. }
\end{figure}

There is a clear $K^{*0} (\bar{K}^{*0})$ signal in the $K \pi$
invariant mass spectrum for $\psip \to \gamma K^+ K^- \pi^+ \pi^-$
candidates. The $\psipto\gamma K^{*0}K^-\pi^+ +c.c.$ and $\gamma
K^{*0}\bar K^{*0}$ branching fractions are measured. The
$\psipto\gamma K^{*0}K^-\pi^+ +c.c.$ branching fraction includes the
contribution from the $\psip \to \gamma \kstaro \kstarb$, and the
$\psip\to \gamma K^{*0}K\pi$ detection efficiency includes the
effect of this contribution. Table \ref{Tot-nev} summarizes the
branching fractions or upper limits for the $\psip$ radiative decays
analyzed. We also report the differential branching fractions of
$\psip$ decays into $\gamma\ppbar$, $\gamma 2(\pipi)$, $\gamma K^+
K^- \pipi$, and $\gamma \kskp$, as shown in Fig.~\ref{diffbr}.

\begin{table}
\begin{center}
\caption{\label{Tot-nev} Results for $\psipto\gamma +hadrons$. For
each final state, the following quantities are given: the number of
events for $m_{hs}<2.9\GC$ in $\psip$ data, $N^{Tot}$; the number of
background events from $\psip$ decays and QED processes, $N^{Bg}$;
the number of signal events,  $N^{Sig}$; and the weighted averaged
efficiency, $\epsilon$; the branching fraction with statistical and
systematic errors or the upper limit on the branching fraction at
the 90\% C.L. Possible interference effects for the
modes with intermediate states are ignored.}
\begin{tabular}{cccccc} \hline \hline
Mode & $N^{Tot}$ & $N^{Bg}$ & $N^{Sig}$ & $\epsilon$(\%) &
$\BR(\times 10^{-5})$\\\hline
$\gamma p\bar{p}$ & $329$ & $187$ & $142\pm18$ & 35.3 & 2.9$\pm$0.4$\pm$0.4 \\
$\gamma 2(\pi^+\pi^-)$ & $1697$ & $1114$ & $583\pm41$ & 10.4  & 39.6$\pm$2.8$\pm$5.0\\
$\gamma K^0_S K^+\pi^-+c.c.$  & $-$ & $-$ & $115\pm16$ & 4.83 & 25.6$\pm$3.6$\pm$3.6 \\
$\gamma K^+ K^-\pi^+\pi^-$ &$361$ &$229$  &$132\pm19$ & 4.94   & 19.1$\pm$2.7$\pm$4.3 \\
$\gamma K^{*0} K^+\pi^-+c.c.$&$-$ &$-$ & $237\pm39$ & 6.86 & 37.0$\pm$6.1$\pm$7.2\\
$\gamma K^{*0}\bar K^{*0}$&$58$&$17$&$41\pm8$&2.75& 24.0$\pm$4.5$\pm$5.0\\
$\gpppr$& $55$ & $38$ & $17\pm7$ &4.47 & 2.8$\pm$1.2$\pm$0.7 \\
$\gkkkk$ & $15$ & $8$  & $<14$ & 2.93& $<4.0$\\
$\gamma3(\pp)$& $118$ & $95$ & $<45$& 1.97 & $<17$\\
$\gamma2(\pi^+\pi^-)K^+K^-$&$17$ & $13$ & $<15.5$ & 0.69 & $<22$ \\
\hline \hline
\end {tabular}
\end{center}
\end{table}

\begin{figure}[ht]
\centerline{\hbox{ \psfig{file=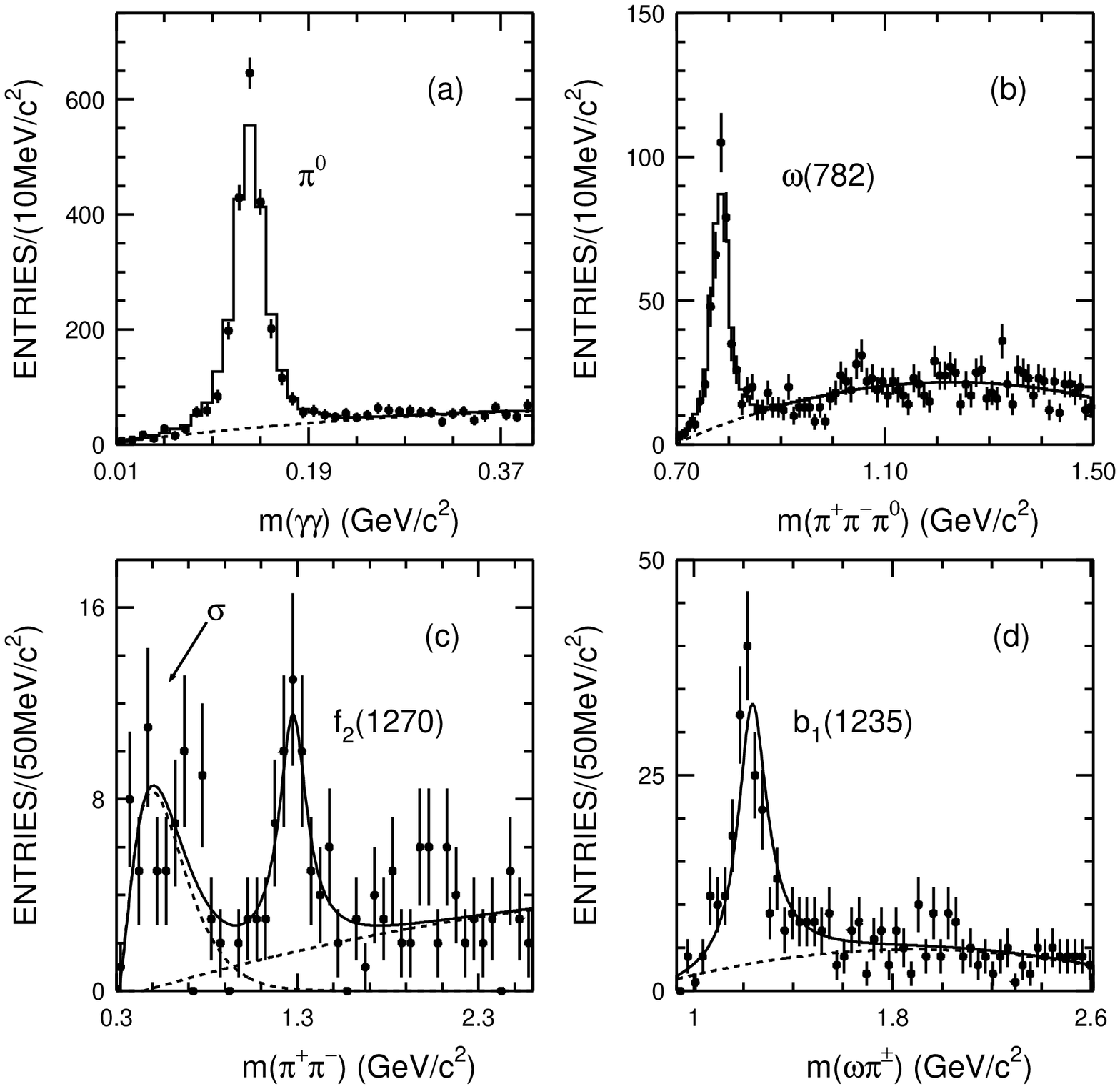,width=0.5\textwidth,
height=8.0cm}}} \caption{\label{m5piall} Invariant mass
distributions with fits for $\psi(2S)\to\pi^02(\pi^+\pi^-)$, where
dots with error bars are data; the solid histograms and curves
denote the fit results. (a) $\gamma\gamma$; (b) $\pi^+\pi^-\pi^0$
with $|m_{\gamma\gamma}-0.135|<0.03\GC$; (c) $\pi^+\pi^-$ with
$|m_{\pi^+\pi^-\pi^0}-0.782|<0.05\GC$; and (d) $\omega\pi^\pm$
with $\omega f_2(1270)$ events removed for the $\psipto\pi^0
2(\pipi)$ candidate events. Resonance parameters are fixed to
their world averaged values~\cite{PDG}.}
\end{figure}

For $\psip$ decays into $\pi^0 2(\pipi)$ and $\pi^0\fptk$, the event
selections are similar to those for $\psip\to\gamma2(\pipi)$ and
$\gamma\fptk$, respectively, but two photons are required. The
numbers of signal events are obtained by fitting the photon pair
invariant mass distributions, and the results are listed in
Table~\ref{Br-pi0bg}. For $\psipto\pi^0 2(\pipi)$ candidate events,
intermediate resonances including $\sigma~[f_0(600)]$, $f_2(1270)$,
$\omega$, and $b_1(1235)$ are observed in the invariant mass
distributions of two pions, three pions, and $\omega\pi$, as shown
in Fig.~\ref{m5piall}. The results for these resonances are given in
Table~\ref{Br-pi0bg}, together with the world averaged
values~\cite{PDG}, and $Q_h~[=\BR(\psip\to h)/ \BR(J/\psi\to h)]$.
Our measurement of ${\cal B}[\psi(2S)\to \omega f_2(1270)]$
     agrees with the previous measurement using the same data
     sample; thus it cross checks the previous result~\cite{VT}.

Table~\ref{syserr} lists the sources of the systematic errors on
the branching fractions. The systematic error caused by MDC
tracking and the kinematic fit is estimated by using simulations
with different MDC wire resolutions~\cite{SIMBES}. The systematic
errors on photon and charged particle identification are taken as
2\% per photon~\cite{SIMBES} and 2\% per charged
particle~\cite{SIMBES}, respectively. The difference of the fit to
the $\chi^2_{4C}$ distribution between MC simulation and data for
$\psip\to \g \chi_{c0}$, $\chi_{c0}\to hadrons$ is about 3\%, which
is taken as the systematic error of the $\chi^2$ fit method. The
uncertainty of the total number of $\psip$ events is
4\%~\cite{npsp_moxh}, the uncertainty of the background estimation
varies from 1-25\% depending on the channel and background level,
and the uncertainties of the branching fractions used are taken
from Ref.~\cite{PDG}. Adding up all these sources in quadrature,
the total systematic errors range from 7 to 28\% depending on the
channel.

In Fig.~\ref{diffbr}, broad peaks appear in the $m_{p\bar p}$ and
$m_{4\pi}$ distributions at masses 1.9-2.5~GeV/$c^2$ and
1.4-2.2~GeV/$c^2$, respectively, which are similar to those
observed in $J/\psi$ decays into the same final
states~\cite{jpsi-gpp,g4pi}. The possible structure within these
broad peaks cannot be resolved with our samples. No obvious structure
is observed in the other final states with the current
statistics. The branching fractions below
$m_{hs}<2.9~\textrm{GeV}/c^2$ in this Letter sum up to
0.26\%~\cite{note1} of the total $\psip$ decay width, which is
about a quarter of the total expected radiative $\psip$ decays.
This indicates that a larger data sample is needed to search for
more decay modes and to resolve the substructure of the $\psip$
radiative decays.

In summary, we  report first measurements of the branching fractions
of $\psip$ decays into $\gamma\ppbar$, $\gamma 2(\pipi)$, $\gamma
\kskp$, $\gamma K^+ K^- \pipi$, $\gamma\kstaro K^- \pi^+ +c.c.$,
$\gamma K^{*0}\bar K^{*0}$, $\gamma\pipi\ppbar$, $\g2(\kk)$,
$\gamma3(\pp)$, and $\gamma\tppkk$, and the differential branching
fractions for $\psip$ decays into $\gamma\ppbar$, $\gamma 2(\pipi)$,
$\gamma K^+ K^- \pipi$, and $\gamma \kskp$ with $m_{hs}$ less than
$2.9\GC$. The branching fractions for $\psip$ decays into
$\pi^0\fptk$ are measured for the first time. The measurements of
$\psip$ decays into $\piz2(\pp)$, $\omega\pp$, and $b_1^\pm\pi^\mp$
are consistent with the recent measurements by the CLEO
collaboration~\cite{CLEO-95} and previous measurements~\cite{PDG}.


\begin{table}
\begin{center}
\caption{\label{Br-pi0bg} Results of $\psipto\pi^0 +hadrons$. Here
$N^{Sig}$ is the number of signal events, $\epsilon$ is the
detection efficiency, $\BR$ is the measured branching fraction,
$\BR^{\textrm{PDG}}$ is the world averaged value, and $Q_h$ is
defined in the text.}
\begin{tabular}{cccccccc} \hline \hline
Mode: $h$  & $N^{Sig}$  & $\epsilon$(\%)  & $\BR(\times 10^{-4})$
& $\BR^{\textrm{PDG}} (\times 10^{-4})$& $Q_h$(\%)\\
\hline
$ \piz2(\pp)$   & $2173\pm53$ & $6.32$ & $24.9\pm0.7\pm3.6$&$23.7\pm 2.6$&$10.5\pm2.0$\\
$\omega\pp$     & $386\pm23$  & $3.74$ & $8.4\pm0.5\pm1.2$&$6.6\pm1.7$&$11.7\pm2.4$\\
$\omega f_2(1270)$ & $57\pm13$& $3.65$ & $2.3\pm0.5\pm0.4$ &$2.0\pm0.6$&$5.4\pm0.6$\\
$b_1^\pm\pi^\mp$& $202\pm21$  & $3.24$ & $5.1\pm0.6\pm0.8$ &$3.6\pm0.6$&$17.0\pm4.2$ \\
$\pi^0\fptk$  & $65\pm17$  & $0.46$ & $10.0\pm2.5\pm1.8$  &---&---\\
 \hline \hline
\end {tabular}
\end{center}
\end{table}

\begin{table}[htbp]
\begin{center}
\caption{\label{syserr}{ Summary of the systematic errors. }}
\begin{tabular}{lc}
 \hline
      Source & Uncertainty     \\\hline
Wire resolution & 5-14\% \\
Photon detection & 2\%/photon \\
Particle identification & 2\%/track \\
Signal fit & 3\% \\
Background estimation & 1-25\% \\
Number of $\psip$ & 4\% \\
Intermediate states & 1-3\% \\\hline
 Total & 7-28\% \\\hline
\end {tabular}
\end{center}
\end{table}

The BES collaboration thanks the staff of BEPC and computing center
for their hard efforts. This work is supported in part by the
National Natural Science Foundation of China under contracts Nos.
10491300, 10225524, 10225525, 10425523, 10625524, 10521003,
10225522, the Chinese Academy of Sciences under contract No. KJ
95T-03, the 100 Talents Program of CAS under Contract Nos. U-11,
U-24, U-25, the Knowledge Innovation Project of CAS under Contract
Nos. U-602, U-34, and the Department of Energy under Contract No.
DE-FG02-04ER41291 (U. Hawaii).


%

\begin{thebibliography}{99}
\bibitem{massres} The $p\bar p$ mass resolution in
the fitted region is less than 3 $\textrm{MeV}/c^2$ and neglected
in the fit.
\end{thebibliography}
\end{document}